\documentclass[%
superscriptaddress,
amsmath,amssymb,
aps, physrev,
prb,
floatfix,
twocolumn,
showkeys,
showpacs,
]{revtex4-2}

\usepackage{float}
\usepackage{soul}
\usepackage[dvipsnames]{xcolor} 
\usepackage{amsmath,amssymb,bm}
\usepackage{graphicx}
\usepackage{dcolumn}
\usepackage{braket}
\usepackage[colorlinks=true,citecolor=blue,urlcolor=blue,linkcolor=blue]{hyperref}
\hypersetup{
pdftitle={hBN},
pdfsubject={hBN},
pdfauthor={Mehdi Arfaoui},
pdfkeywords={hBN}
}

\usepackage[section]{placeins}
\usepackage{comment}

\newcommand{\LSc}[1]{\textcolor{blue}{\hl{#1}}}
\newcommand{\LSa}[1]{\textcolor{blue}{#1}}

\begin{document}
\title{Stacking, symmetry and exciton fine structure in boron nitride bilayers}

\author{Mehdi Arfaoui}
\email{arfaouimehdi20@yahoo.fr}
\affiliation{%
Université Paris-Saclay, ONERA, CNRS, Laboratoire d'étude des microstructures (LEM), F-92322 Châtillon Cedex, France.
}%
\affiliation{%
 Université Paris-Saclay, CEA, CNRS, SPEC, 91191 Gif-sur-Yvette, France.
}%

\author{Sylvain Latil}
\affiliation{%
 Université Paris-Saclay, CEA, CNRS, SPEC, 91191 Gif-sur-Yvette, France.
}%


\author{Lorenzo Sponza}%
\affiliation{%
Université Paris-Saclay, ONERA, CNRS, Laboratoire d'étude des microstructures (LEM), F-92322 Châtillon Cedex, France.
}

\date{\today}%

\begin{abstract}
Engineering of two-dimensional material heterostructures offers a broad range of degrees of freedom through the selection of type and number of layers and their relative stacking configurations. In these compound materials, the stacking sequence plays a central role in determining electronic, mechanical, and optical properties.
Here, we present a theoretical study of its signature on the excitons of hexagonal boron nitride bilayers based on first-principle calculations within the GW approximation and the Bethe-Salpeter equation.
We explain how stacking-dependent symmetry variations lead to distinct electronic structures, particularly near the band edge around the K high-symmetry point, and consequently to markedly different excitonic landscapes. Indeed, we analyze the symmetry-governed character of the low-lying excitonic states, identifying their bright or dark nature for each stacking configuration and performing a systematic comparison. Our results clarify how van der Waals interlayer coupling reshapes excitons in hexagonal boron nitride bilayers and outline prospects for twist-controlled exciton engineering in this material.
\end{abstract}


\maketitle

\section{Introduction}

Excitons dominate the optical response of two-dimensional (2D) materials due to reduced dielectric screening and enhanced electron-hole (\textit{e-h}) Coulomb interactions~\cite{Wang2018RMP}
that lead to large binding energies~ \cite{Chernikov2014PRL,He2014PRL,Ugeda2014NatMater,ferreira2019excitons} and thus thermal stability up to room temperature.
Thanks to this, tweaking excitonic properties in 2D materials is way less challenging than in the bulk and has been demonstrated by dielectric environment engineering~ \cite{Stier2016NanoLett, Peimyoo2020ACSAMI}, application of mechanical strain~ \cite{Long2020Nanoscale} or external fields~\cite{Unuchek2018}.
These achievements have placed 2D materials to the front of the stage for a broad range of optical applications 
such as quantum light source~\cite{Chakraborty2019Nanophotonics, Gale2026} and light emitting, detecting and optoelectronic devices ~\cite{Fu2021LaserPhotonicsRev,Shang2018SMTD,Wang2018RMP}.


In this respect, sp$^2$ boron nitride (hBN) is a particularly appealing material in virtue of its qualities as single photon emitter~\cite{Shaik2021,Carbone2025} and for the exceptional characteristics of its excitons~\cite{Cudazzo2016, Cassabois2016,Galvani2016,Schue_2019,Roux2021PRB}. 
In the bulk form, hBN is a stack of monolayers \--- often referred to as AA$^\prime$ stacking \--- and has a wide bandgap of at least $6$ eV. 
Like in graphene and graphite, bondings have a $\sigma$ hybridization component that develops parallel to the atomic planes and a $\pi$ component that extends some \AA{} above and below the atomic planes and stems from the hybridization of the B and N $p_z$ orbitals. 
Concerning optical properties, the reported experimental lifetime of its fundamental exciton is about $27$ ns~\cite{Roux2021PRB} with a quantum yield of nearly $50\%$~\cite{Schue_2019} which makes it a remarkably strong emitter in the deep ultraviolet~\cite{Watanabe2004,Schue_2019, Cassabois2016,science.aeb2095} despite its indirect band gap.
The fundamental exciton develops essentially in plane and has an extraordinary small planar extension \--- the average electron-hole distance being about 7~\AA{}~\cite{Paleari2018,Roux2021PRB}. 
Most of these characteristics are preserved when reducing the dimensionality to few layers or down to the monolayer (ML)~\cite{Galvani2016,Cudazzo2016}.
On the other hand, excitons in hBN are pretty sensitive to the local environment owing to their compactness and the protrusion out of plane of the $\pi$ orbitals, thus they can be used to probe characteristics like the number of layers or the stacking sequence.
In fact, theoretical works have predicted changes resulting from differences in the local dielectric screening~\cite{Bourrellier2014,Paleari2018, Darshana2018,Sponza2018PRB_polymorphs} or in the coupling with phonons~\cite{Zanfrognini2023}.
Luminescence experiments have demonstrate with success that it is possible to discern between the AA$^\prime$ and the Bernal, or AB, stacking in the bulk~\cite{Zanfrognini2023}, as well as the number of layers in moderately thick slabs~\cite{schue2016dimensionality}.

A particularly relevant field where local stacking drives the device physics  is twistronics~\cite{Tecorralco2026}. 
Recently, it has been demonstrated that, by simply twisting two thin slabs of hBN on top of each other, it is possible to exalt or suppress specific emission channels~\cite{Lee2021a,Roux_PhysRevX,science.aeb2095}, or tune the characteristics of quantum emitters embedded in one of the two slabs~\cite{Gale2026}.
There are theoretical and experimental evidences that upon twist \--- especially at small angles \--- massive reconstructions occur which lead to the formation of large triangular domains~\cite{Kenji2021,Moore2021,Woods2021,Walet2021}.
Within each domain, layers are perfectly aligned according to specific and stable stacking sequences that vary from one domain to the other in an alternated fashion. 
If some conditions are met, this may even give rise to ferroelectric domains~\cite{Kenji2021,Moore2021,Woods2021}.
To investigate the optical properties of twisted hBN structures at small angles, it is essential to gain insight on how each domain determines separately the properties of the fundamental exciton.

A prototypical system to investigate the impact of the stacking registry on excitonic properties is the hBN bilayer (BL). 
Previous theoretical works on the subject have primarily focused on the evolution of the bandstructure and gap character~\cite{constantinescu2013stacking, WenJiangPRB106, Alexander1602, PhysRevMaterials.9.024004, Gilbert_2019} but have seldom discussed excitonic properties.
To our knowledge, the few works tackling hBN excitons from this perspective~\cite{Henriques2022, Bourrellier2014, Zhu2024, Paleari2018, Sponza2018PRB_polymorphs} are either restricted to specific stackings (\textit{e.g.}, AA$^\prime$ or AB) or to twisted geometries at angles barely relevant for experiments or are carried out in the bulk. 
Here, we focus on excitons in the optical limit in hBN BLs across the high symmetry stacking configurations encountered in small-twist-angle domains. 
Our intent is that of scrutinizing the relation between stacking registry, symmetry, and a series of excitonic properties comprising the spectral fine structure, selection rules and localization in real space.

The article is organized as follows.
In Sec.~\ref{sec2}, we introduce the stacking landscape of hBN BLs and analyze the associated symmetry properties in the atomic and electronic structure for each configuration.
In Sec.~\ref{sec3}, we present a detailed study on the excitonic properties listed above.
Conclusions are drawn in Sec.~\ref{summary}.
Additional computational and technical details are provided in the Appendices.

\section{Stacking Landscape and Symmetry}\label{sec2}

\subsection{Atomic structure}
\begin{figure}
  \centering
  \includegraphics[width=0.98\linewidth]{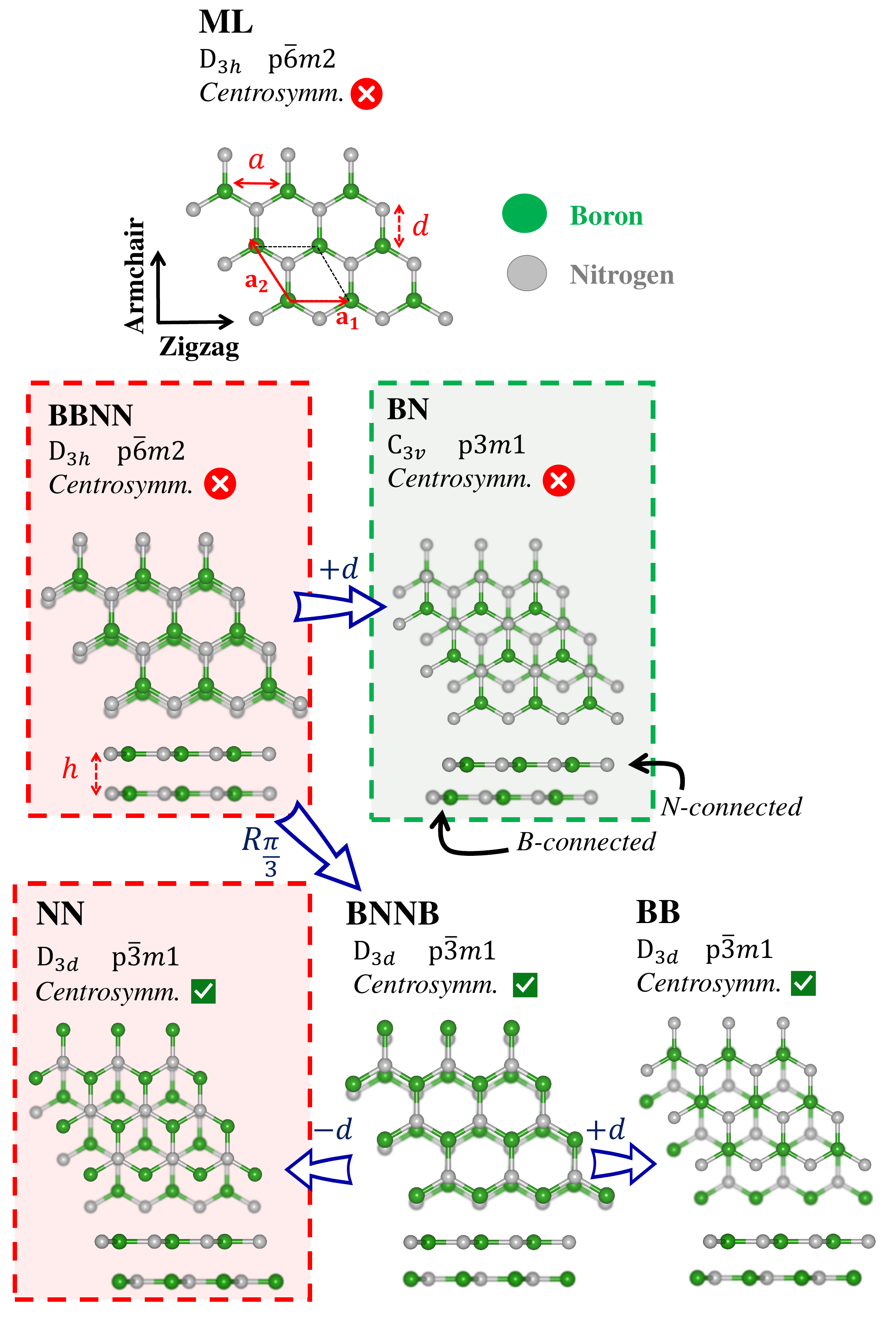}
  \caption{Top: hBN ML with boron (green) and nitrogen (grey) atoms in the 2D honeycomb lattice. Middle and bottom: Symmetry-related stacking configurations in hBN BL generated by in-plane translations and rotations. Each panel corresponds to a distinct operation applied to the reference stacking. The green dashed boxes represent the most energetically stable stackings, while those enclosed by red dashed boxes correspond to less stable configurations. The in-plane crystallographic directions are identified as zigzag and armchair, defined with respect to the Cartesian axes $(x,y)$. The in-plane primitive lattice vectors denoted by $(\mathbf{a}_1, \mathbf{a}_2)$ are indicated with red arrows. The dashed line represents the primitive unit cell.}
  \label{fig:STRUCTURE}
\end{figure}

\begin{figure}
  \centering
  \includegraphics[width=\linewidth]{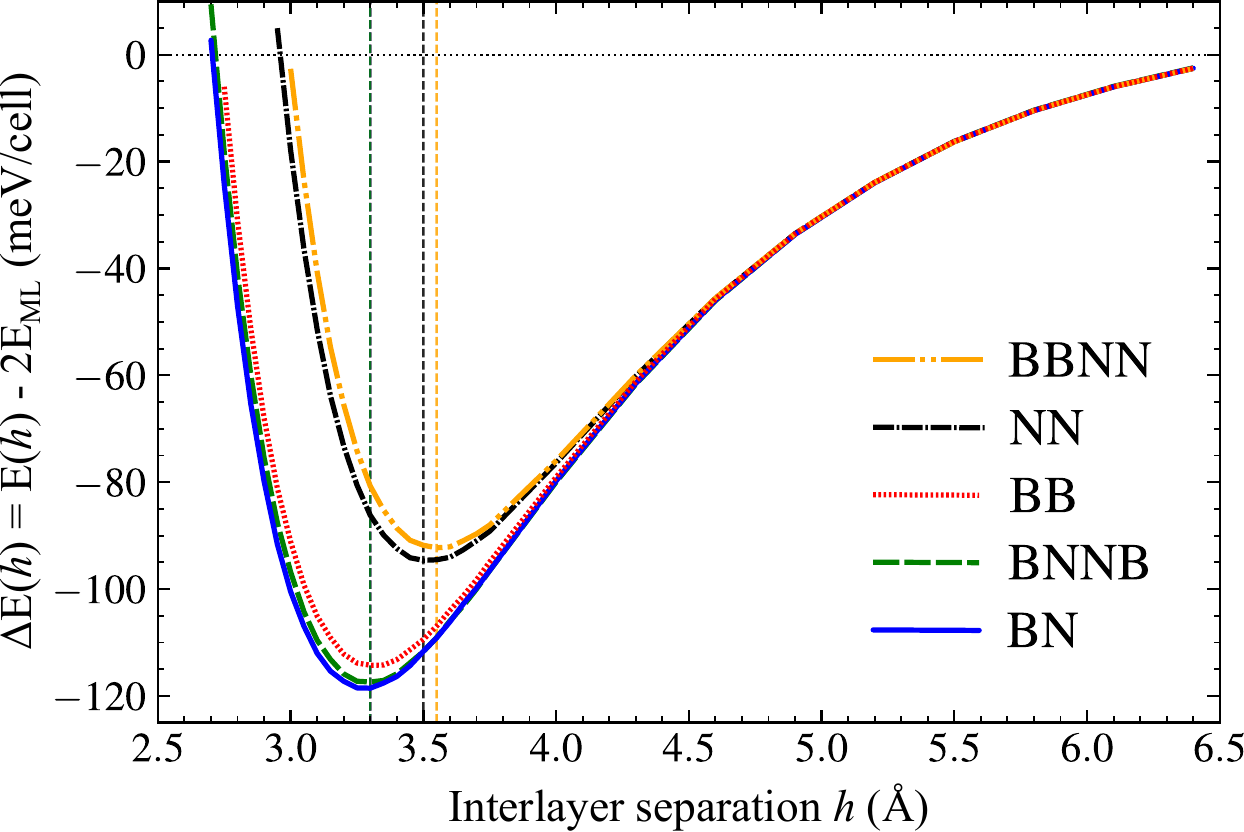}
  \caption{
   Interlayer binding energy $\Delta E(h)=E_{\text{BL}}(h)-2E_{\text{ML}}$  per unit cell as a function of the interlayer separation $h$ for the different stackings.
   $E_{\mathrm{ML}}$  is the total energy/unit cell of an isolated ML. The vertical dotted lines indicate the equilibrium separations $h$.}
  \label{fig:STABILITY}
\end{figure}

Fig.~\ref{fig:STRUCTURE} illustrates the atomic structure and crystallographic orientation of hBN ML and BL. The planar lattice consists of alternating Boron and Nitrogen atoms forming a honeycomb motif.  
The in-plane B-N bond length is denoted by $d = a/\sqrt{3}\sim 1.44$ \,\AA{}, with $a$ the in-plane lattice constant.
 For the BL system, we define five distinct high-symmetry stacking registries: 
following Latil et al.~\cite{latil2023structural} we adopt the notation BBNN (often referred to as AA stacking), BNNB (usually called AA'), BN (Bernal or AB stacking), BB, and NN.
To highlight the transformations linking them, we start their construction from the ML.
By simply duplicating the ML along $z$ with no additional transformation to neither of the two layers, one obtains the BBNN stacking.
Holding fixed the bottom layer of the latter, while translating the top layer by $+ d$ along the armchair direction, the BN stacking configuration is generated.
An equivalent NB stacking (sometimes designated as BA) can be generated by a translation of $- d$, but since these two configurations are equivalent under a swapping of all N and B atoms, only the BN stacking is considered in the following.
We now go back to the BBNN stacking and, while keeping the lower layer in place, we rotate the top layer by $60^{\circ}$ about the center of a hexagon ($\mathcal{R}_{\pi/3}$ transformation), or swap N with B atoms in the top layer which is equivalent. 
We end up with the BNNB stacking.
From this one, by shifting the top layer by $+ d$ or $- d$ along the armchair direction, one obtains either the BB or the NN stacking.

These different stacking registries endow the BL with different symmetries.
The less symmetric stacking is the BBNN which is non-centrosymmetric and belongs to the D$_{3h}$ point group (layer group p$\bar{6}$m2), exactly as the ML. 
The most symmetric ones are the BNNB, BB and NN which are all centrosymmetric and belong to the point group D$_{3d}$ (layer group p$\bar{3}$m1)~\cite{latil2023structural}.
The BN stacking lies in between. Its symmetry reduces to the C$_{3v}$ point group (layer group p3m1) and also lacks of inversion symmetry. 
Unlike the other four stackings, BN features symmetry-inequivalent layers, which we denote as \textit{N-connected} and \textit{B-connected} according to whether N or B atoms are vertically aligned with atoms of the opposite layer. This interlayer asymmetry, combined with the polarity of B-N bonds, gives rise to an out-of-plane electric polarization, underlying the experimentally observed stacking-dependent ferroelectricity in bilayer hBN~\cite{Kenji2021,Woods2021}. Reversing the stacking order (BN to NB) reverses the polarization and exchanges layer-localised properties~\cite{chadiza2025first,Kenji2021} (see for example Sec.~\ref{sec:wfc}).

Given the different vertical alignment, the repulsion between the layers is different and thus their average vertical separation. 
In Fig.~\ref{fig:STABILITY}, the equilibrium interlayer separation  exhibits sizeable variations among the different stacking configurations.
For the energetically favorable stackings, the equilibrium interlayer distances are approximately $h^{\text{BN}} \approx h^{\text{BNNB}} \approx h^{\text{BB}} \approx 3.3$\,\AA.
In contrast, the less stable stackings display larger equilibrium separations, specifically $h^{\text{NN}} = 3.5$\,\AA{} and $h^{\text{BBNN}} = 3.55$\,\AA.
This systematic correlation between stability and interlayer separation is consistent with previous reports~\cite{latil2023structural, Peres2011}. The computational details of the structural relaxations are given in Appendix~\ref{COMPUTE}

\subsection{Electronic structure}

\begin{figure}
  \centering
  \includegraphics[width=0.98\linewidth]{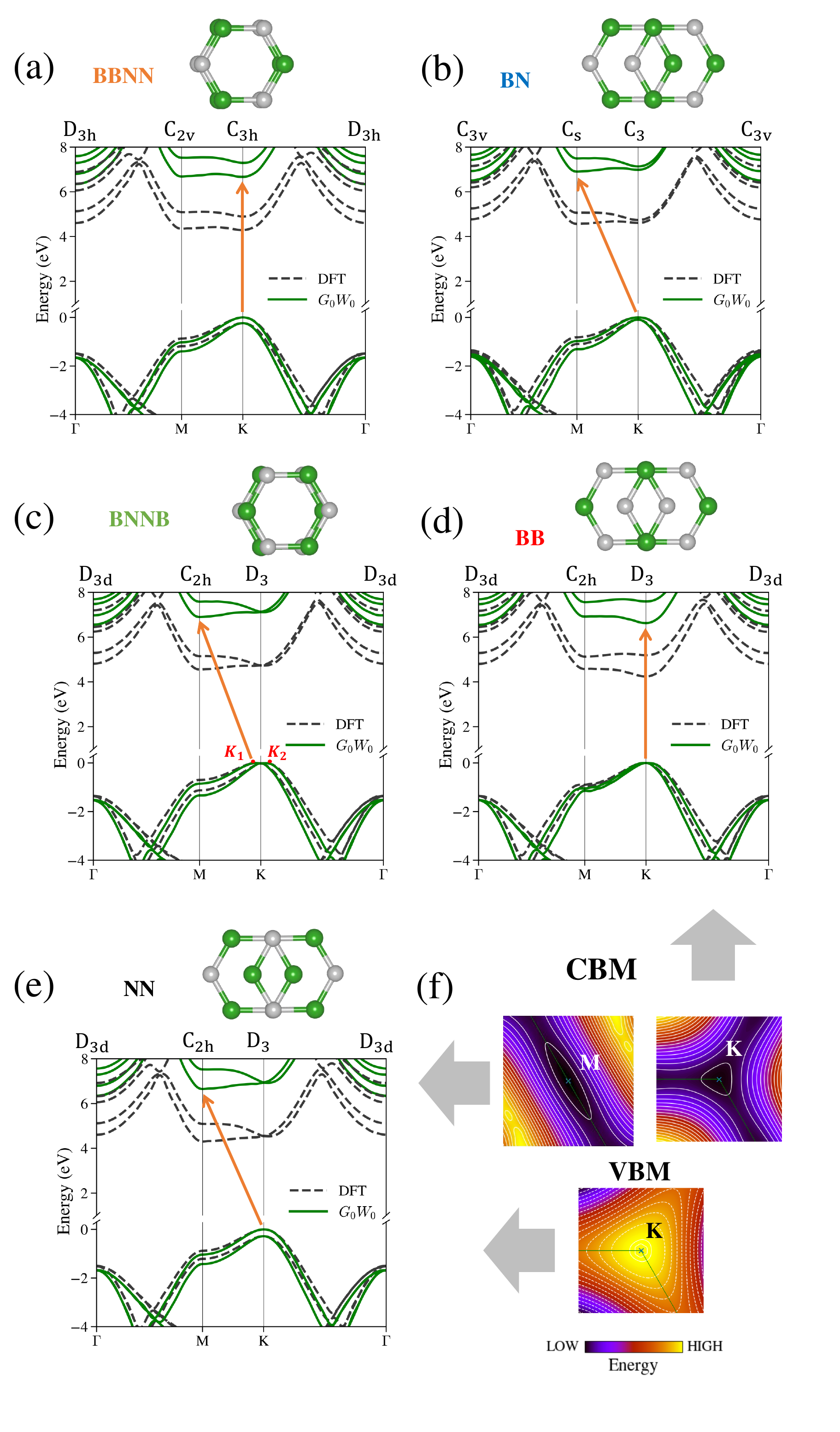}
\caption{\textbf{Quasiparticle electronic structure of bilayer hBN across five stacking configurations.} (a)-(e) DFT (black dashed) and $G_0W_0$ (green solid) band structures along the $\Gamma$MK$\Gamma$ path for BBNN, BN, BNNB, BB, and NN, respectively. Atomic configurations are shown above each panel, with the little groups indicated at the high-symmetry points. Orange arrows mark the quasiparticle band gaps. (f) Representative constant-energy contours around the CBM at M and K (top) and the VBM at K (bottom). The color scale indicates increasing energy from dark to bright.}
  \label{fig:qpGW}
\end{figure}

We now focus on the impact of the stacking on the electronic states relevant for excitonic properties. 
In Fig.~\ref{fig:qpGW}, we report the DFT and G$_0$W$_0$ band structure of the five stackings together with the little groups associated to each high-symmetry point.
All the computational details of DFT and G$_0$W$_0$ are provided in Appendix.~\ref{COMPUTE}.
Quasiparticle corrections open the band gap widely over $\sim 2$ eV and slightly modify the dispersion, mainly in the vicinity of the $K$ valley and more severely in empty states.
In agreement with the literature, we found that the electronic structure depends on the stacking configuration~\cite{Gilbert_2019, mengle2019impact}.
The valence band maximum (VBM) is at K in four stackings out of five, the only exception being the BNNB configuration where it is at $K_1=(3/10,\,1/3,\,0)$, close to K.
A bit more delicate is the identification of the conduction band minimum (CBM).
From a numerical perspective, the actual quasiparticle CBM is in $\Gamma$ in all our calculations but this is an artifact due to a delicate convergence issue discussed in Appendix~\ref{NFE}. 
Since these states are utterly irrelevant for optical properties, we will continue the analysis as if they were absent from the calculation.
In this sense, a higher variability is observed for the (physical) CBM that is either in $M$ or in $K$ depending on the system.
We summarize in Table~\ref{kEGSPLIT} the direct  and indirect  quasiparticle band gaps of the five hBN BLs.
Our present results are in excellent agreement with previously reported theoretical calculation~\cite{mengle2019impact, latil2023structural}.

\renewcommand{\arraystretch}{1.6}
\begin{table}
\centering
\caption{Band gap (eV) with band edges (VBM/CBM), band splittings (meV) at M and K in the top valence and bottom conduction bands and binding energy of the $B_1$ exciton (eV). The symbol -- indicates degeneracy due to a band crossing.}
\begin{tabular}{lc|cc|cc||c}
& & \multicolumn{2}{c|}{valence}
& \multicolumn{2}{c||}{conduction} & \\
Stacking & band gap & $M$ & $K$ & $M$ & $K$ & $E_{gap}^{QP}-E_X^{B_1}$ \\
\hline
BBNN  & 6.576 ($K/K$) & 386 & 301 & 887 & 739 & 1.439\\
BN    & 6.852 ($K/M$) & 359 & 143 & 602 & 194 & 1.714 \\
BNNB  & 6.846 ($K_1/M$) & 455 & --  & 626 & -- & 1.803\\
BB    & 6.535 ($K/K$) & 41  & --  & 608 & 1051 & 1.498 \\
NN    & 6.594 ($K/M$) & 354 & 305 & 837 & -- &  1.706\\
\hline
\end{tabular}\label{kEGSPLIT}
\end{table}

Besides the nature and the width of the gap, other characteristics are impacted by the symmetry of the stacking registry.
The electronic structure of the BBNN can be easily understood if seen as resulting from the interaction of two perfectly identical MLs. 
Each state of the BBNN bandstructure is either a bonding or an antibonding combination of the corresponding states of the two MLs. For instance, 
at K the generic state $\psi_{nK}$ is  classified by its eigenvalue $s_h=\pm 1$ under reflection $\hat{\sigma}_h$ on a plane passing between the two layers, namely $\hat{\sigma}_h\psi_{nK}=s_h\psi_{nK}$. This classification is demonstrated by the signed wavefunctions $\text{Re}\,\psi_{nK}(\mathbf r)$ shown in Fig.~\ref{fig:B_AB_BBNN} of the appendix \ref{BBNN_sym}.
This interaction leads to the avoided band crossings observed in BBNN, when the interacting states have the same symmetry.
In the other high-symmetry stackings this interpretation is still valid in its essence, but their different symmetry group imposes other constraints, for instance enforcing band crossing at K either in valence (BB), or in conduction (NN) or in both (BNNB). The representative constant-energy contours in Fig.~\ref{fig:qpGW}(f) illustrate the band-edge dispersion near K and M. At K, both the valence and the conduction band contours exhibit trigonal warping, consistent with the underlying threefold rotational symmetry $C_3$. In contrast, the approximately elliptical contours near M reveal an anisotropic conduction-band dispersion.

To be more quantitative, we report in Table~\ref{kEGSPLIT} the band splitting in $K$ and $M$ between the two highest valence bands and the two lowest conduction bands.
Splittings are systematically larger in the conduction band than in the valence bands as a consequence of the stronger B-B interaction~\cite{Sponza2024}, but besides this, they show a pronounced variability between stackings and k-points that is another manifestation of how the stacking modifies the interlayer hybridization.
Effective masses extracted from the quasiparticle band structure can be found in Appendix~\ref{EFFMASS}.

\begin{figure}
  \centering
  \includegraphics[width=\linewidth]{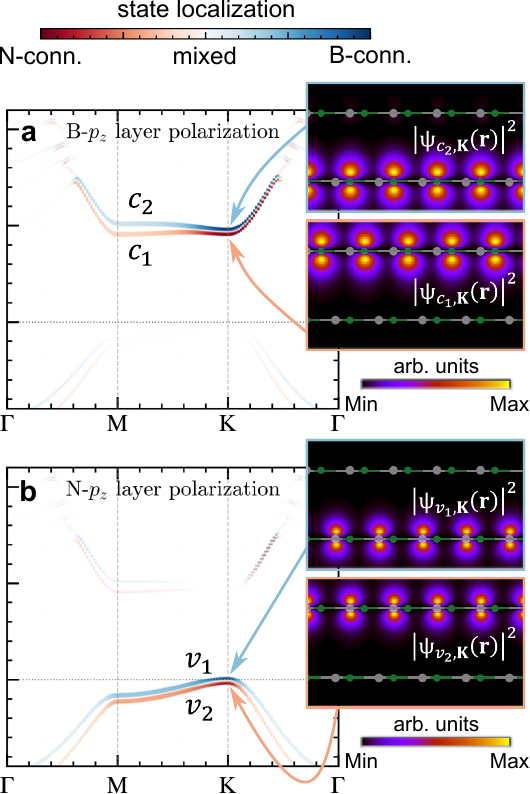}
   \caption{\textbf{Layer-resolved electronic character of the BN stacking.} (a) B-$p_z$ and (b) N-$p_z$ layer polarization, $L_{p_z}^{B}(\mathbf{k},E)$ and $L_{p_z}^{N}(\mathbf{k},E)$, along $\Gamma$MK$\Gamma$. Blue and red indicate dominant contributions from the \textit{B-connected} and \textit{N-connected} layers, respectively, while the color opacity reflects the projected spectral weight. Insets show the side-view real-space probability densities $|\psi_{nK}(\mathbf{r})|^2$ of the conduction ($c_1,c_2$) and valence ($v_1,v_2$) states at K.}
  \label{fig:pz_orbital_BN}
\end{figure}

We stress that also the BN stacking does not display any band crossing.
In this respect its band structure looks similar to that of BBNN, but for a very different reason.
While in BBNN two identical MLs mix and form symmetric and antisymmetric replicas of the ML bandstructure, in the BN this interpretation does not hold as the two layers are inequivalent.
Its band structure must be actually interpreted as resulting from two weakly interacting MLs, each feeling a slightly different environment. 
The lowest energy ``copy'' stems from the \textit{N-connected} layer, while the highest energy one comes from the \textit{B-connected} (respectively top and bottom layer in our convention). 
Actually, for a more quantitative description, a certain amount of mixing between the two layers should be taken into account, but this simplified picture is a powerful reading key.
This is clearly illustrated in Fig.~\ref{fig:pz_orbital_BN}, where the band structure is colored according to the layer-resolved $p_z$ polarization, defined as
$L_{p_z}^{s}(\mathbf{k},E) = \frac{ I_{s,p_z}^{\text{N-conn}}(\mathbf{k},E) - I_{s,p_z}^{\text{B-conn}}(\mathbf{k},E) }{ I_{s,p_z}^{\text{N-conn}}(\mathbf{k},E) + I_{s,p_z}^{\text{B-conn}}(\mathbf{k},E) }$.
Here, $I_{s,p_z}^{\ell}(\mathbf{k},E)$ denotes
the $k$-resolved, $p_z$-projected spectral
intensity of species $s=\mathrm{B,N}$ on
layer $\ell=\mathrm{B\text{-}conn,N\text{-}conn}$.
Accordingly, $L_{p_z}^{s}=+1$ ($-1$)
indicates localization on the\textit{N-connected}
(\textit{B-connected}) layer, whereas
$L_{p_z}^{s}=0$ corresponds to equal
contributions from both layers.
Consistently, the real-space probability
densities at K [Fig.~\ref{fig:pz_orbital_BN}(b)]
show that the states within each pair,
$(c_1,c_2)$ and $(v_1,v_2)$, are
predominantly localized on opposite layers.

\section{Excitonic properties} \label{sec3}

State-of-the-art ab initio calculations of excitonic properties are carried out by solving the Bethe Salpeter equation (BSE) using G$_0$W$_0$ quasiparticle energies~\cite{Rohlfing2000,Onida2002}.
The solution of the BSE relies on the diagonalization of a static excitonic hamiltonian written on the basis of independent-particle transitions.
It yields the excitonic eigenvalue $E_X^{\lambda}$ and eigenvector $\mathbf{A}^{\lambda}$ for the excitonic state $\lambda$.
The exciton spectrum in the optical limit (\emph{direct} excitons) is $\varepsilon_2 \propto \sum_{\lambda} f_\lambda \delta(E_X^{\lambda} - \hbar\omega)$ and is obtained as a sum over the excitonic resonances $E_X^{\lambda}$, each weighted by its oscillator strength $f_\lambda \propto \left| \sum_{vc\mathbf{k}} A^{\lambda}_{vc\mathbf{k}} \,\mathbf{e}\cdot\mathbf{M}_{cv\mathbf{k}} \right|^2$
where $\mathbf{M}_{cv\mathbf{k}} =\int d\mathbf{r}\; \psi^{*}_{c\mathbf{k}}(\mathbf{r})\, \hat{\mathbf{p}}\, \psi_{v\mathbf{k}}(\mathbf{r})$ is the interband optical matrix element in the velocity gauge.
Here, $\hat{\mathbf{p}}=-i\hbar\nabla$, $\mathbf{e}$ is the polarization vector of the incident light, and  $\psi_{n\mathbf{k}}(\mathbf{r})$ is the single electron Bloch wave function of band index $n$ and momentum $\hbar \mathbf{k}$  calculated in the real-space coordinate $\mathbf{r}$.

All parameters of the following calculations are detailed in Appendix~\ref{COMPUTE}.

\subsection{Absorption spectra}\label{EXC_SIG}

\begin{figure*}
  \centering
  \includegraphics[width=1.\textwidth]{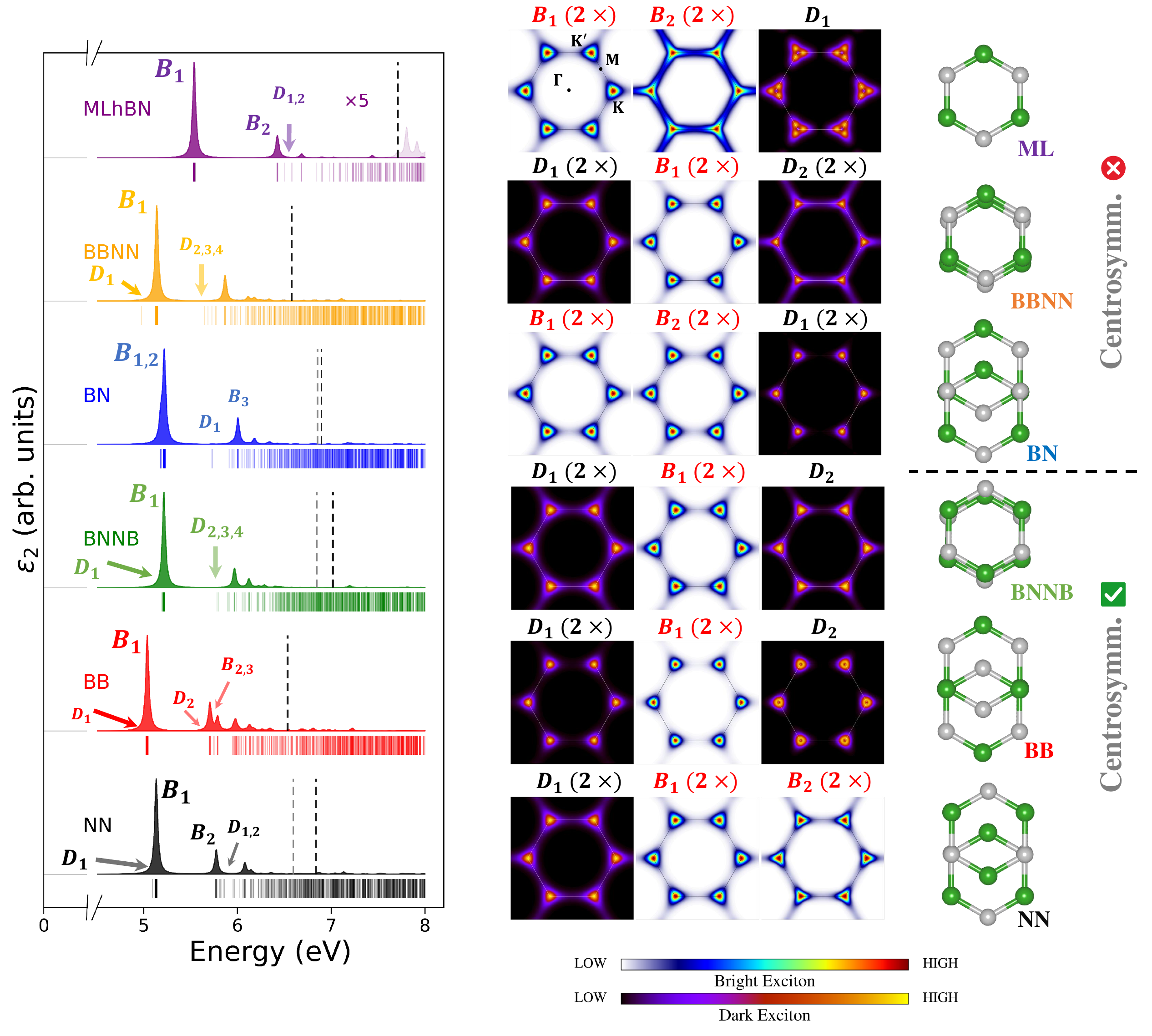}
  \caption{\textbf{Excitonic optical spectra and $\mathbf{k}$-resolved wavefunctions in hBN ML and BLs.} 
  The left panels present $\varepsilon_2$ computed within $G_0W_0$+BSE (dark contrast) and $G_0W_0$+IPA (light contrast).
Vertical sticks below the spectra mark the exciton energies $E_X^\lambda$, with their intensity proportional to $f_\lambda$.
Arrows and labels highlight the lowest excitonic states with $B$ labeling bright states and $D$ dark ones.
  The direct and indirect quasiparticle gaps are indicated by vertical dashed black and gray lines respectively.
  We use a broadening of $30$ meV for the optical spectra.
  The right panels display the three lowest excitons' $w_{\mathcal{G}}(\mathbf{k})$.
 We use distinct colormaps for bright and dark states to enhance visibility.
}
  \label{fig:OPTICALSPECTRA}
\end{figure*}

In Fig.~\ref{fig:OPTICALSPECTRA}, the left panel shows the excitonic absorption spectrum of the ML and the BLs.
Spectra correspond to in-plane polarized light (\textit{i.e.}, $\hat{\mathbf{e}}\perp \hat{z}$), noting that the optical response is isotropic within the layer plane.
The independent particle approximation (IPA) spectrum evaluated using G$_0$W$_0$ quasiparticle energies is also shown only for the ML and it is multiplied by factor of $5$ for sake of visibility.
The overall IPA spectrum is roughly 15 times lower in intensity than the excitonic features in the considered energy window and
this ratio is verified in all cases, highlighting the strong signature of excitons in hBN materials.
The right panels show the $\mathbf{k}$-resolved map of the squared amplitude of the exciton envelope wave function, namely $w_{\mathcal{G}}(\mathbf{k}) = \sum_{\lambda \in \mathcal{G}}\sum_{v,c}\left|A_{vc\mathbf{k}}^{\lambda}\right|^2$ for the first three lowest-lying \textit{direct} excitons, with $\mathcal{G}$ the degenerate set corresponding to exciton state $\lambda$. 

In the ML, the response is dominated by an intense sharp peak labeled B$_1$. This stems from the first exciton which is a two-fold degenerate $1s$-like bright exciton of $E'$ symmetry with a large binding energy $E_b = E_{gap}^{QP}-E_X^{B_1} \sim 2.1$~eV, in good agreement with previous works~\cite{Galvani2016, PhysRevB.106.045118}.
Upon stacking, the landscape becomes richer owing to the increase and densification of excitonic states.
Although the $B_1$ peak dominates in most stackings, $B_2$ becomes dominant in the BN configuration, where $B_1$ has a reduced oscillator strength. Additional dark and bright excitonic states appear within a narrow energy window, with stacking-dependent spectral shifts. 
The $\mathbf{k}$-resolved maps show that $B_1$ remains confined to the pockets around $K$ and $K^{\prime}$ in all stackings, clearly reflecting the threefold rotation symmetry $C_3$.
Like in the ML, it has a $1s$-like orbital character in all BLs.
In Table~\mbox{\ref{kEGSPLIT}}, we report the calculated exciton binding energies of $B_1$ in the five BLs.
Remarkably, in all BLs the $B_1$ resonance falls at an energy between $1.4$ and $1.8$~eV lower then that of the ML $B_1$ despite the higher binding energy of the latter.
This is consistent with the screening being weaker in the ML than in the BLs. This leads to larger exciton binding energies and wider band gap opening than in the BLs, with the latter effect being stronger than the former.
A detailed ranking of low-energy excitons in each system can be found in Fig.~\ref{fig:BSEEGLEVELS} of Appendix~\ref{APPEXC}.

In the follwoing, we will first discuss the excitonic properties of the BBNN, BNNB, BB and NN stackings, and leave  the distinct optical response of the BN stacking for the end.
In these four stackings the second bright exciton $B_2$ lies approximately $\sim 0.5-0.9$~eV above $B_1$.
In the ML, this has been assigned to a $2p$-like excitonic state with $E^\prime$ symmetry~\cite{PhysRevB.106.045118, Galvani2016}.
Although a similar spectral feature is observed in these BL stackings, their $B_2$ states exhibit distinct k-resolved distributions. We therefore do not assign them solely from their energy separation and defer their detailed characterization to the wavefunction analysis.

In centrosymmetric configurations (BNNB, BB and NN), the excitons can be classified according to their even ($g$) or odd ($u$) parity under inversion. The lowest excitonic manifold contains a dark $D_1$ state and a bright $B_1$ state. Their energy separation is compatible with a Davydov splitting (DS) arising from the coupling between equivalent layers. We define its magnitude as $\Delta_{BD} = E_X^{B_1} - E_X^{D_1}$.
DS has been predicted theoretically in multilayer hBN~\cite{Paleari2018} and observed experimentally in multilayer MoSe$_2$, where it was used to estimate the number of layers~\cite{kim2016davydov}. In the present BLs, $\Delta_{BD}$ provides an effective measure of how interlayer coupling modifies the lowest excitonic pair.
Here, we obtain $\Delta_{BD}\approx 8$ meV for BB, $20$ meV for BNNB, and $40$ meV for NN. This ordering does not follow the interlayer distance alone, but reflects the full excitonic coupling, including exchange, and possible charge-transfer (CT) mixing (cfr. Sec.~\ref{sec4}). 
A related dark-bright pair is also found in the noncentrosymmetric BBNN stacking. Although inversion symmetry is absent, the horizontal mirror plane $\sigma_h$ exchanges the two equivalent layers. The $D_1$ and $B_1$ manifolds are both doubly degenerate and are separated by $\Delta_{BD}\approx 164$ meV, the largest value among the investigated stackings.

The symmetry origin of all these dark-bright pairs is examined in Sec.~\ref{SELRULE}, while their common excitonic parentage is investigated in Sec.~\ref{sec:wfc}.

Once again, the BN stacking stands out among the five BL configurations as its low-energy absorption spectrum displays a completely different fine-structure. 
Its lowest exciton is bright, featuring a shouldered peak that arises from two nearly overlapping bright excitons $B_1$ and $B_2$, separated by about $38$~meV.
Because the two layers are inequivalent and both excitons are optically active, this pair is not assigned to the symmetry-selected dark-bright DS scenario discussed above.
Instead, following the idea expressed before, we interpret its spectrum as the superposition of two MLs inserted in slightly different environments. In this perspective, both $B_1$ and $B_2$ are replicas of the ML $B_1$ exciton, each being localised on one of the two inequivalent layers. The absorption spectrum alone does not provide enough information to determine in which layer they are preferentially localized. We need to investigate the selection rules to disclose from which layer the two excitons originate.

\subsection{Onset exciton selection rules} \label{SELRULE}

\begin{figure}
  \centering
  \includegraphics[width=1.\linewidth
   ]{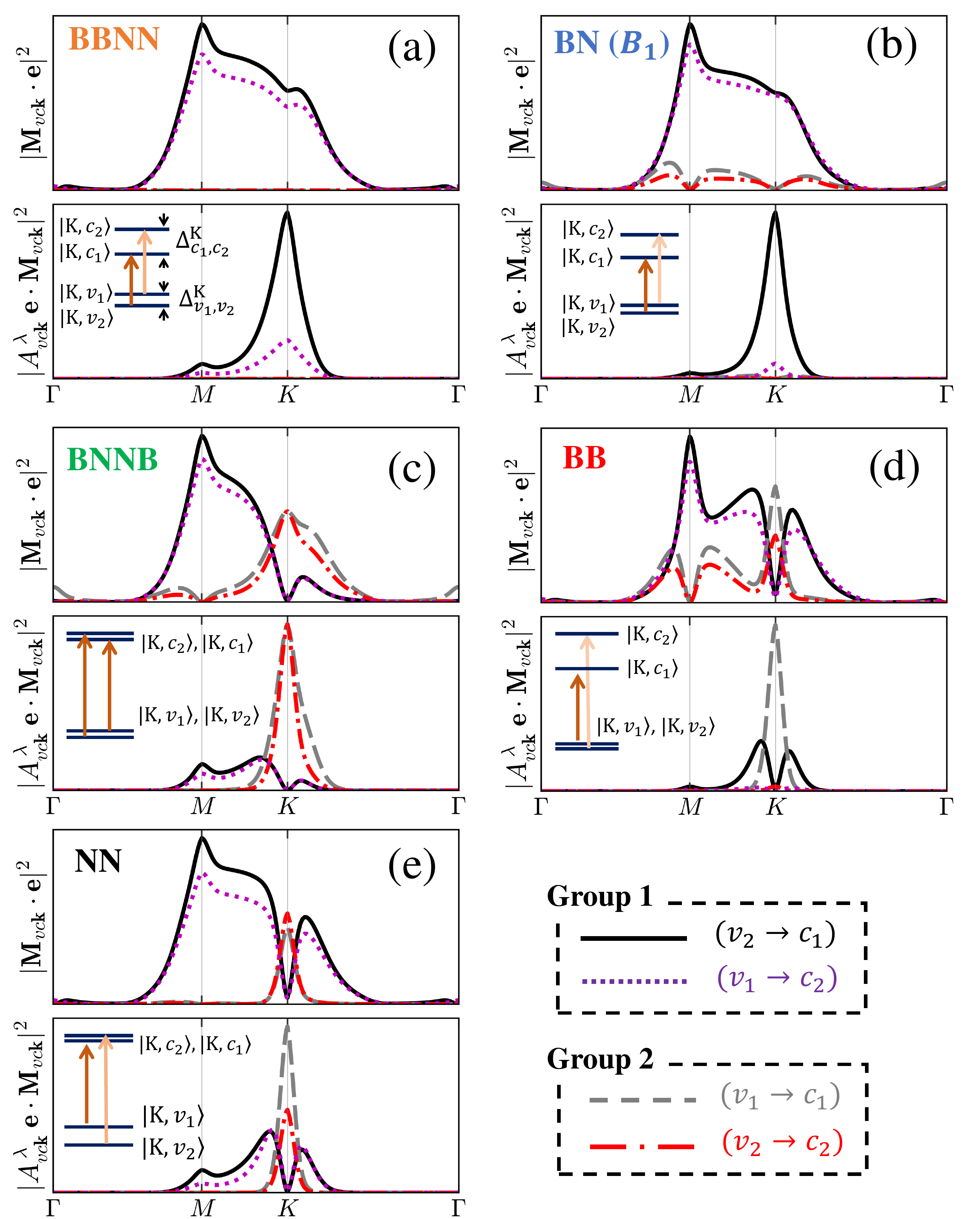}
  \caption{Momentum-resolved band contributions to the IPA optical dipole matrix elements (top panel of each stacking configuration) and to the excitonic oscillator strength of the first bright exciton $B_1$ (bottom panel) plotted along the high symmetry path $\Gamma$MK$\Gamma$ for (a) BBNN, (b) BN, (c) BNNB, (d) BB, and (e) NN stackings.
  For the excitonic contribution, we sum the weight over the degenerate states belonging to the $B_1$  exciton.
  Insets in bottom panels: illustration of band edge predominant transition at $K$ point contribution to $B_1$, the opacity of the arrows is proportional to $f_\lambda$. We note that each plot is normalized with respect to its own maximum value.
  }
  \label{fig:BANDCONTRIBUTION_IPA_Acvk}
\end{figure}

We now examine the symmetry origin of the lowest dark-bright exciton pairs and the band-resolved optical selection rules along the $\Gamma$MK$\Gamma$ path.
In BLs with the $D_{3d}$ symmetry, the in-plane dipole operator transforms as $E_u$. Consequently, the lowest $E_u$ exciton is optically allowed ($B_1$), whereas its $E_g$ partner $D_1$ is dark. In the $D_{3h}$-symmetric BBNN stacking, the in-plane dipole transforms as $E^{\prime}$, making the $E^{\prime}$ exciton bright ($B_1$) and its $E^{\prime\prime}$ partner dark ($D_1$). Each of these excitons is doubly degenerate. This information supports the identification of $D_1$ and $B_1$ as DS pair.
The BN case is different. Its $C_{3v}$ symmetry does not exchange the two inequivalent layers, and the in-plane dipole transforms as $E$. Its two lowest $E$ excitons, $B_1$ and $B_2$, are therefore both optically allowed.
$B_1$ is the lower-energy bright exciton, while $B_2$ lies about $38$~meV higher and has a substantially larger oscillator strength.

To investigate the microscopic origin of the $B_1$, we consider the two highest valence bands, $v_1$ and $v_2$, and the two lowest conduction bands, $c_1$ and $c_2$, where the lower label identifies states closer to the gap. The band-to-band transitions of these states provide the dominant dipole contributions to this exciton.
The top panels of Fig.~\ref{fig:BANDCONTRIBUTION_IPA_Acvk} show the bare interband dipole strengths $|M_{vc\mathbf{k}}\cdot \mathbf{e}|^2$ for each stacking.
In all stackings, the dipole strength is mainly activated along the MK segment, with some weak contribution near the $\Gamma$ point.
The four relevant transitions can be classified into two groups: $v_1 \rightarrow c_2$ and $v_2 \rightarrow c_1$ form the first group, while $v_1 \rightarrow c_1$ and $v_2 \rightarrow c_2$ form the second group.
The first group dominates the bare dipole strength over most of the M-K segment with stronger contributions at M. Transitions occurring at K are not dominant, and they are even dipping to zero in the BNNB, BB and NN stackings.
The second group is overall weaker and displays an opposite trend, with vanishing contributions at M and peaked ones at K.
In this respect, it is worth stressing the singularity of the BBNN stacking, that is the only one for which transitions of the second group are rigorously dipole forbidden.
This can be understood by recalling that the signed wavefunctions $\text{Re}\, \psi_{n,K}(\mathbf{r})$ shown in Fig.~\ref{fig:B_AB_BBNN} classify $v_2$ and $c_1$ as mirror-even, bonding states with $s_h=+1$, and $v_1$ and $c_2$ as mirror-odd, antibonding states with $s_h=-1$.
Since the in-plane dipole operator is even under the horizontal mirror operation $\sigma_h$, transitions between states with opposite mirror parity are dipole forbidden, whereas transitions between states with the same mirror parity are allowed.

However, in the actual spectrum what matters is the exciton oscillator strength $|A^{B_1}_{vc\mathbf k}\: \hat{\mathbf{e}} \cdot \mathbf{M}_{vc\mathbf{k}}|^2$ where dipole matrix elements are weighted by the exciton wave function, here reported in the bottom panels of Fig.~\ref{fig:BANDCONTRIBUTION_IPA_Acvk}.
Taking this into account dramatically changes the landscape.
In all stackings, the excitonic weighting strongly suppresses the contributions near M, while those near K dominate. This said, the relative contributions of the four band-to-band channels depend on the stacking.
In the BBNN and BN stackings, the first group channels remain dominant \--- with a way stronger contribution from the $(v_2 \rightarrow c_1)$ component \--- while the second group channels are suppressed.
For BBNN, this directly follows from the $\sigma_h$ selection rule described above. For BN, the different weights of the available channels instead reflect the inequivalence of its two layers.
In fact, when compared with Fig.~\ref{fig:pz_orbital_BN}, these results show that $B_1$ is essentially an in-plane exciton localized on the N-connected (top) layer. The higher energy exciton $B_2$ instead is predominantly associated with the $(v_1 \rightarrow c_2)$ transition and is localized on the B-connected (bottom) layer. 
In contrast, in BB, NN and BNNB stackings, bands are degenerate at $K$ either in valence, in conduction or both. In these systems, the weight of the ($v_1 \rightarrow c_1$) and the ($v_2 \rightarrow c_2$) channels is anti-correlated with the energy of the corresponding IP transition, i.e. higher energies lead to lower contributions and equal energies lead to equal contributions (the BNNB case).

To summarize these results, the schematics in Fig.~\ref{fig:BANDCONTRIBUTION_IPA_Acvk} illustrate the band-resolved selection rules at $K$ for $B_1$, with arrow opacity indicating the relative optical contribution of each transition. In Sec.~\ref{sec:wfc}, we compare their momentum- and real-space distributions to determine whether they also share a common excitonic origin, as expected for Davydov partners.

\subsection{Onset exciton wave function}
\label{sec:wfc}

\begin{figure}
  \centering
  \includegraphics[width=1.\linewidth]{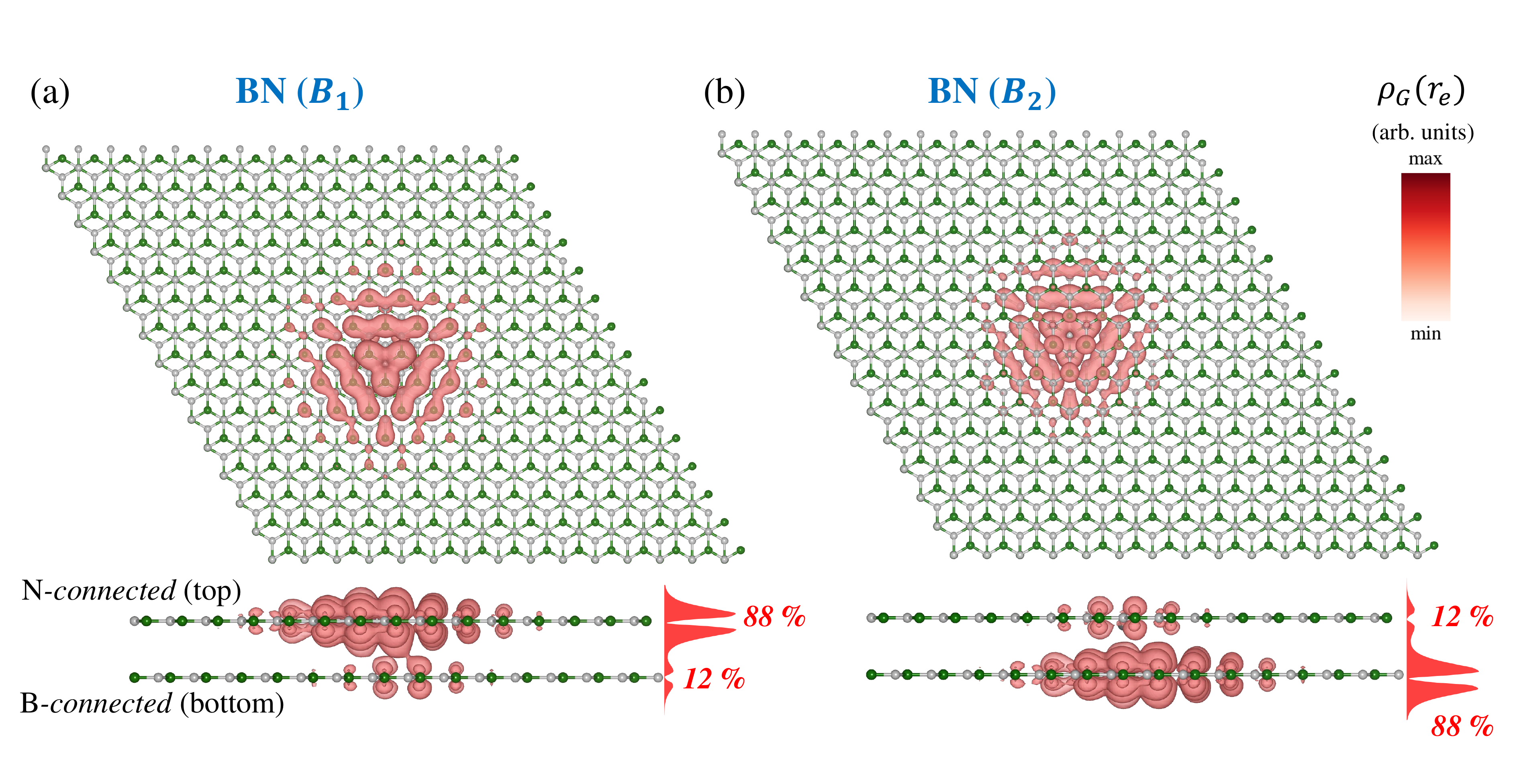}
  \caption{Real-space excitonic wave function of the lowest bright (a) $B_1$ and (b) $B_2$ exciton in BN stacking.}
  \label{fig:exciton_real_space_wf_BN}
\end{figure}

To gain more insight into the nature of the lowest excitons, we now analyze their spatial character by inspecting the real-space exciton wave function $\Psi^\lambda(\mathbf{r}_e,\mathbf{r}_h) = \sum_{v,c,\mathbf{k}} A^{\lambda}_{vc\mathbf{k}}\; \psi_{c\mathbf{k}}(\mathbf{r}_e)\; \psi^{*}_{v\mathbf{k}}(\mathbf{r}_h)$, where $\mathbf{r}_e$ and $\mathbf{r}_h$ are the electron and hole coordinates, respectively.
To visualize the real-space wave function, a common practice is to fix the hole position and plot $|\Psi^\lambda(\mathbf{r}_e,\mathbf{r}_h)|^2$ as a function of $\mathbf{r}_e$.
Actually, to reduce the dependence on the arbitrariness of the hole position, 
in this work we identify a set of relevant hole positions $\{\mathbf{r}_h^i\}_i$ such that their sum reflects the symmetries of the system~\mbox{\cite{Paleari2018}} and we average over them.
We place the hole $0.5$ \AA\; above and below each of the two N atoms of the unit cell, for a total of four positions per system, and compute
\begin{equation}
\begin{split}
\rho_\mathcal{G}(\mathbf{r}_e) &\propto \frac{1}{4} \sum_{i} \rho_\mathcal{G}^i(\mathbf{r}_e)
\quad\text{with} \\ 
\rho_\mathcal{G}^i(\mathbf{r}_e) &= \sum_{\lambda \in \mathcal{G}} \left| \Psi^\lambda(\mathbf{r}_e , \mathbf{r}_h^i ) \right|^2\,.
\end{split}
\label{eq:four_hole_density}
\end{equation}
The corresponding plots are reported in Figure~\ref{fig:exciton_real_space_wf} for the $B_1$ exciton of the BNNB, BBNN, BB and NN stackings, and in Figure~\ref{fig:exciton_real_space_wf_BN} for the $B_1$ and $B_2$ excitons of the BN stacking.

\begin{figure*}
  \centering
  \includegraphics[width=0.9\textwidth]{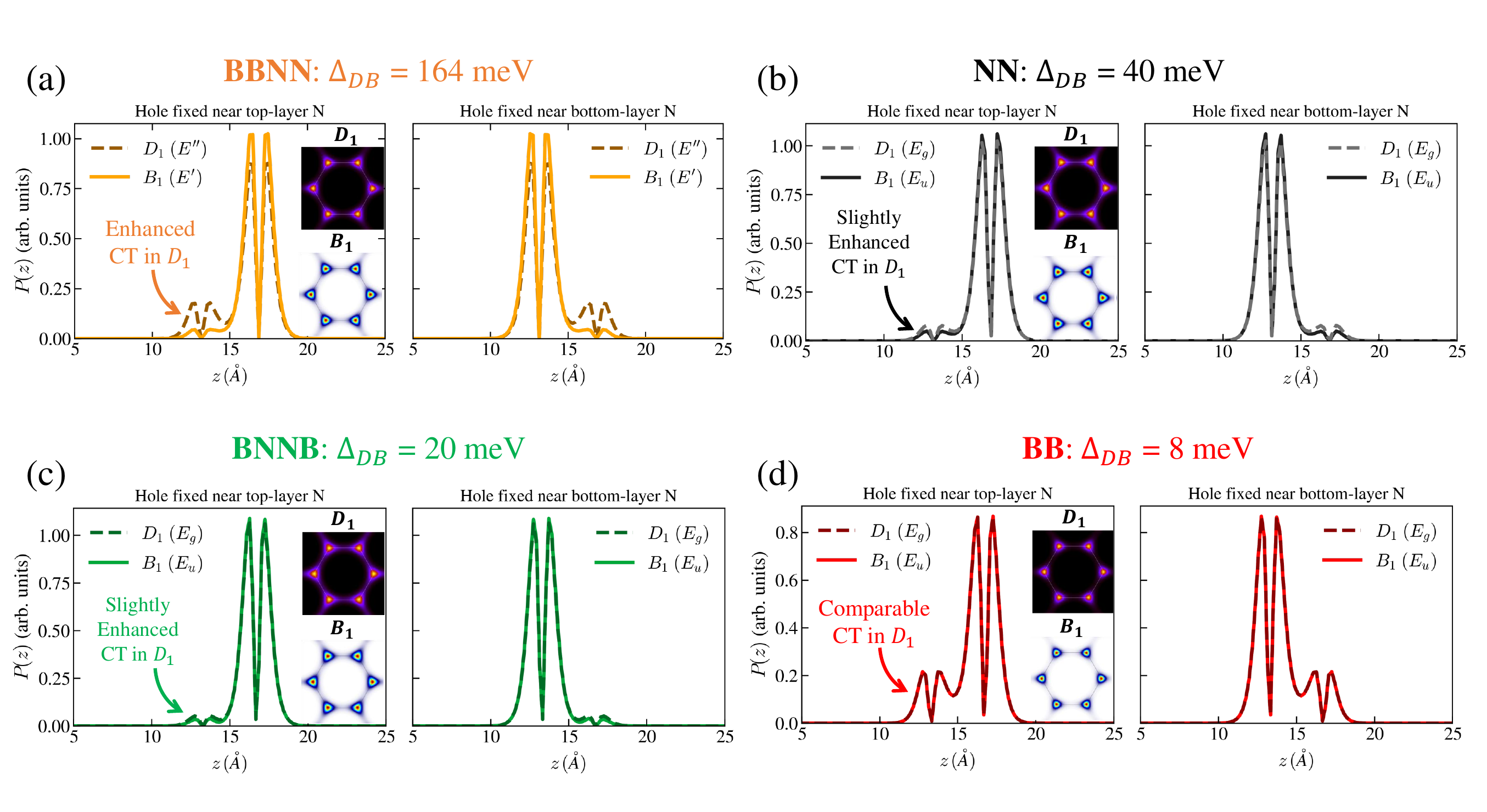}
  \caption{Electron-density profiles along the out-of-plane direction, $P_{\mathcal{G}}^{i}(z)$, of the lowest $D_1$ (dashed lines) and $B_1$ (solid lines) excitons in (a) BBNN, (b) NN, (c) BNNB, and (d) BB stackings with the hole fixed near an N atom of either the top or bottom layer.}
  \label{fig:fixed_hole_CT_DD}
\end{figure*}

To quantify the exciton spatial extension, we calculate the mean \textit{e-h} separation $\big\langle d_{eh}\big\rangle_{\mathcal G}(\mathbf r_h^0)
\propto
\dfrac{\int d^3r_e\;
\rho_{\mathcal G}(\mathbf r_e;\mathbf r_h^0)\;
\big|\mathbf r_e-\mathbf r_h^0\big|}
{\int d^3r_e\;
\rho_{\mathcal G}(\mathbf r_e;\mathbf r_h^0)}$. 
Figures~\ref{fig:exciton_real_space_wf} and \ref{fig:exciton_real_space_wf_BN} show that, irrespective of the stacking, the spatial extension of the B$_1$ exciton has roughly a radius of $\sim3$ unit cells
(i.e. $\big\langle d_{eh}\big\rangle_{\mathcal G}\sim 8$ \AA).
The small spatial extension is consistent with strong \textit{e-h} Coulomb attraction.
In this respect, the $B_1$ exciton in BLs does not display features  strikingly dependent on the stacking sequence, nor notably different  from thinner or thicker layers, including the bulk~\mbox{\cite{Roux2021PRB, Paleari2018, Galvani2016, Cudazzo2016}}.

We next quantify the electron-density contribution from layer $\ell$ by computing $\%_\ell = \dfrac{\int_{z\in\ell} \rho_{\mathcal{G}}(\mathbf r_e)\,d^3r_e}{\int \rho_{\mathcal{G}}(\mathbf r_e)\,d^3r_e}$ where $z\in \ell$ denotes the spatial region containing layer $\ell$. The nearly equal averaged weights found for BB, BNNB, NN, and BBNN are consistent with their equivalent layer contributions. However, because the density is averaged over symmetry-related hole positions.
In contrast, as clearly visible in Fig.~\ref{fig:exciton_real_space_wf_BN}, the BN stacking exhibits a pronounced layer asymmetry in $B_1$ and $B_2$, consistent with the layer inequivalence stressed in discussing Figs.~\ref{fig:pz_orbital_BN} and~\ref{fig:BANDCONTRIBUTION_IPA_Acvk}.
In fact, $B_1$ is predominantly localized on the \textit{N-connected }(top) layer, with approximately $88$\% of the integrated density, whereas $B_2$ is predominantly localized on the \textit{B-connected} (bottom) layer, also with approximately $88$\%.

The four-hole-averaged electron densities of the $B_1$ exciton for ML hBN and the BBNN, BNNB, BB, and NN stackings are reported in Appendix, in Fig.~\mbox{\ref{fig:exciton_real_space_wf}}.

\subsection{Charge transfer and Davidov splitting}
\label{sec4}

To retain information about the \textit{e-h} layer correlation, we also calculate the out-of-plane probability profile for each fixed-hole position, $P_{\mathcal{G}}^{i}(z) = \int dx\,dy\, \rho_{\mathcal{G}}^{i}(x,y,z)$. 
Unlike the averaged densities, these profiles resolve the electron distribution relative to the layer hosting the fixed hole.
To discuss the DS occuring in bulk hBN, a model has been proposed~\cite{Koskelo2017} where the interlayer hopping induces a mixing of Frenkel-like and charge transfer (CT) interlayer excitons.
In this respect, analysing $P_{\mathcal{G}}^{i}(z)$ allows a quantification of such a mixing and its importance in determining the Davidov splitting of the BBNN, BNNB, BB and NN stackings.
Fig.~\ref{fig:fixed_hole_CT_DD} compares the $P_{\mathcal{G}}^{i}(z)$ profiles of the lowest dark $D_1$ and bright $B_1$ excitons in these four bilayer stackings, with the hole fixed near an N atom in either layer.
For each stacking, $D_1$ and $B_1$ exhibit similar dominant $P_{\mathcal{G}}^{\,i}(z)$ profiles and similar $K/K^{\prime}$ weights. This common momentum- and real-space character shows that they originate from the same parent exciton, further supporting their origin as Davydov pairs.
In both excitons, the dominant peaks of $P_{\mathcal{G}}^{i}(z)$ occur around the same layer as the fixed hole, showing that both states retain dominant intralayer \textit{e-h} correlation. The smaller peaks on the opposite layer indicate a finite CT admixture. In all cases, this is more important for the dark excitons, in agreement with~\cite{Koskelo2017}. In addition, the difference between the CT component of $D_1$ and $B_1$ is directly correlated to the strength of the DS, namely it is the largest in the BBNN ($\Delta_{DB}=164$ meV), the weakest in the BB ($\Delta_{DB}=8$ meV) and in between for the NN ($\Delta_{DB}=40$ meV) and the BNNB ($\Delta_{DB}=20$ meV).

\section{Conclusion}\label{summary}

In this work we have presented a systematic and deep analysis of  how the stacking impacts the electronic and excitonic properties in hBN BLs, with a particular focus on symmetries and a constant comparison with the ML.
Our analysis has been carried out in the framework of state-of-the-art ab initio Many Body Perturbation Theory.

We showed that the five hexagonal stackings of hBN yield distinctive characteristics in the electronic band structures and excitonic properties.
This includes the energy and hierarchy of dark and bright excitons, their degeneracy, their origin in terms of single-particle states and their real-space localization. 
A special attention has been drawn on the singularity of the ``BN'' stacking whose non-equivalence of the layers is at the origin of unique features like the presence of a shoulder in the absorption onset and a peculiar excitonic localization
that can be understood by thinking at the BN stacking as the superposition (and weak mixing) of two MLs inserted in slightly different dielectric environments.

We believe that these results are particularly relevant for the investigation of optical properties of twisted hBN bilayers where massive reconstructions at small twist angles give raise to triangular domains with the most stable stacking configurations among the accessible ones, like the ``BN'' one.

\begin{acknowledgments}
This work has also benefitted from the French National Research Agency grant numbers ANR-21-CE09-0016 and ANR-21-CE09-0025 (respectively the EXCIPLINT and GANESH projects).

\end{acknowledgments}

\section*{Data availability}
The data that support the findings of this study are available from the corresponding author upon reasonable request. In addition, relevant computational details are provided.

\section*{Notes}
The authors declare no competing financial interest.

\appendix

\renewcommand{\thefigure}{S\arabic{figure}}
\setcounter{figure}{0}

\begin{figure*}
  \centering
  \includegraphics[width=0.7\linewidth]{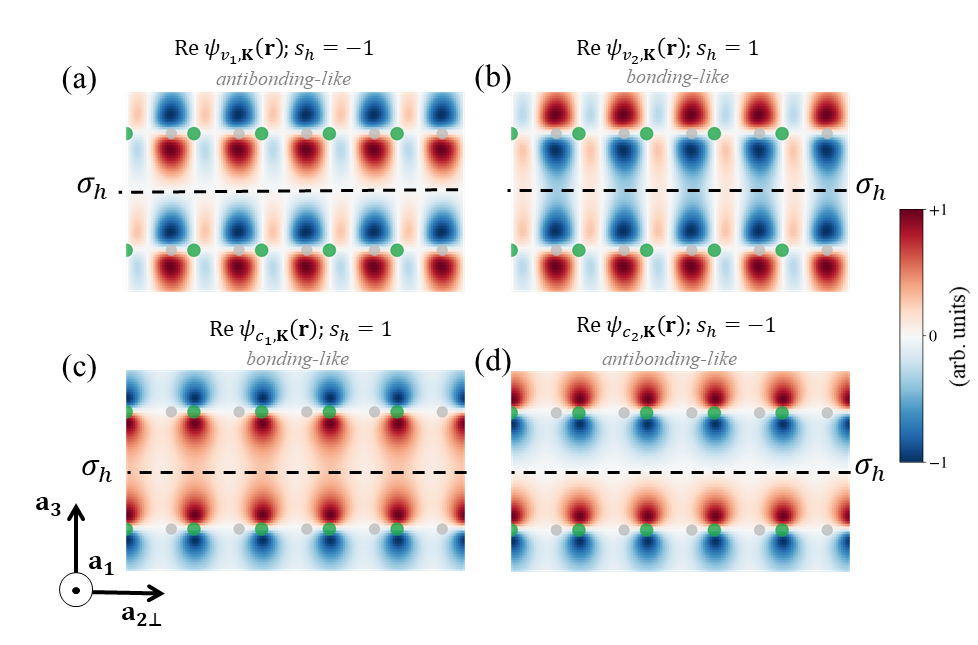}
  \caption{Mirror symmetry of the BBNN at K. Real parts of the wave functions of (a) $v_1$, (b) $v_2$, (c) $c_1$, and (d) $c_2$, shown in a plane perpendicular to the layers. The dashed horizontal line indicates the mirror plane $\sigma_h$. 
  }
  \label{fig:B_AB_BBNN}
\end{figure*}


\section{Computational Details}\label{COMPUTE}

We study the electronic and optical properties of hBN BL using an \textit{ab initio} approach based on  ground-state DFT,  many-body perturbation theory within quasiparticle G$_0$W$_0$ approximation, and then we solve the BSE on top of G$_0$W$_0$  for neutral excitation~\cite{Hybertsen1986,Onida2002,Rohlfing2000}.
\subsection{Density Functional Theory Calculations}
DFT calculations were performed using the plane-wave (PW) pseudopotential method as implemented in the
\textsc{Quantum ESPRESSO} package~\cite{Giannozzi2009,Giannozzi2017,Giannozzi2020}. Exchang-correlation effects were carried out within the generalized gradient approximation using the Perdew-Burke-Ernzerhof  functional~\cite{Perdew1996}. Optimized norm-conserving Vanderbilt pseudopotentials~\LSa{\cite{Hamann2013,dojo}} were employed for both boron and nitrogen atoms. The PW kinetic energy cutoff for the wave functions was set to $70$ Ry, which ensured convergence of total energies and electronic properties. Spin-orbit coupling was systematically neglected due to its negligible effect in hBN. The Brillouin zone was sampled using a $18\times18\times1$ $\mathbf{k}$-point $\Gamma$-centered grid for structural relaxations and self-consistent field (SCF) calculations. Subsequent non-self-consistent field (NSCF) calculations used a denser $30\times30\times1$ $\mathbf{k}$-point mesh and included $600$ electronic bands in order to accurately describe the unoccupied states required for many-body calculations. We set a vacuum of approximately $30$ \AA{} along the out-of-plane direction in order to avoid interactions between periodic images. Structural relaxations were performed using the Broyden-Fletcher-Goldfarb-Shanno algorithm to optimize the in-plane lattice parameter, atomic positions, and interlayer distance until the residual forces on each atom were below than $10^{-4}$ Ry/Bohr, the total energy charge smaller than $10^{-5}$ Ry and the pressure were below $0.1$ kbar.
To properly account for interlayer interactions in BL systems, vdW corrections were included using the semi-empirical Grimme D3 dispersion scheme~\cite{Grimme2010,Grimme2011}. This approach provides a reliable description of long-range dispersion interactions and has been widely used for layered materials. Because the systems investigated are effectively 2D, Coulomb truncation was applied along the out-of-plane direction in order to suppress spurious interactions between periodic replicas~\cite{Sohier2017}.

\subsection{Many-Body Perturbation Theory: G$_0$W$_0$ and BSE}

The GW and BSE calculations were carried out with the YAMBO code~\cite{ Sangalli2019, Marini2009} using a 2D slab cutoff of the Coulomb interaction. The quasiparticle bandstructre were computed within the $G_0W_0$ approximation. The dielectric screening was computed using the plasmon-pole approximation, and a dielectric-matrix cutoff of  $10$ Ry. The GW calculations employed the same $\mathbf{k}$-point sampling used in the NSCF calculations and included $600$ bands to ensure convergence of the dielectric matrix and self-energy $\Sigma = iGW$. Coulomb truncation was applied along the out-of-plane direction to eliminate spurious interactions between periodic replicas.
Excitonic effects in the optical spectra were then included by solving the BSE for the \textit{e-h} two-particle Hamiltonian. Since the focus of this work is on the low-energy absorption region, only the six highest valence bands and the six lowest conduction bands were included in the BSE kernel. The BSE screening was computed using a cutoff of $10$ Ry, while the direct and exchange parts of the BSE kernel were described using cutoffs of $8$ Ry and $45$ Ry respectively. 
The Tamm-Dancoff approximation was adopted, and the optical limit was treated within the dipole approximation. 
For the spatial exciton map plotting, we use a $15 \times 15\times 1$ supercell.

\renewcommand{\arraystretch}{1.6}
\begin{table}[t]
\caption{Electron ($m_e^*$) and hole ($m_h^*$) effective masses (in units of the free-electron mass $m_0$) obtained from parabolic fits around the valence-band maximum and conduction-band minimum of the ML and the different BL stacking configurations.}
\label{tab:effective_mass}
\centering
\begin{tabular}{lcccccc}
\hline
Stacking & Valley & $m_h^*(v_1)$ & $m_h^*(v_2)$ & $m_e^*(c_1)$ & $m_e^*(c_2)$ \\
\hline
ML    & K      & 0.632 & --    & 0.827 & --    \\

BBNN  & K      & 0.669 & 0.581 & 0.761 & 0.878 \\

BN    & K      & 1.350 & 0.445 & 1.085 & 0.889 \\

BNNB  & K$_1$  & 0.706 & --    & 5.173 & --    \\
      & K$_2$  & 0.543 & --    & 0.949 & --    \\

BB    
      & K      & --    & --    & 0.704 & --    \\

NN    & K      & 0.713 & 0.630 & --    & --    \\
      & M      & --    & --    & 0.708 & --    \\

\hline
\end{tabular} 
\end{table}

\section{Conduction Band Minimum and Near-Free-Electron states}
\label{NFE}
At $\Gamma$, the bottom conduction bands have a neat parabolic shape.
These are near-free-electron states which extend over several \AA{}ngstr\"{o}ms above and below the BL~\cite{Paleari_phd,latil2023structural}.
They are extremely hard to converge with the vacuum and this trend is not healed by the Coulomb truncation technique because their occurrence is only partially an artifact of periodic boundary conditions.
Unless including a huge amount of vacuum in the calculation, quasiparticle corrections tend to place them below the K and M valleys so that they constitute the actual numerical bottom conduction in most simulations with hBN surfaces (\textit{cfr.} GW corrections to the ML in~\cite{Galvani2016} or nanotubes~\cite{latil2023structural}).
However, since these states are irrelevant for optical properties, we neglect them completely in our discussion and work as if the CBM were occurring either in M or K depending on the stacking.

\section{Symmetry of BBNN states}\label{BBNN_sym}
Figure~\ref{fig:B_AB_BBNN} shows the mirror symmetry of the BBNN at K. For in-plane polarization, transitions between states with the same mirror eigenvalue are allowed, while those between states with opposite eigenvalues are forbidden. A detailed discussion of the resulting optical selection rules is provided in Sec.~\ref{SELRULE} of the main text.

\section{Effective masses}\label{EFFMASS}
To get more insight into the carrier and exciton dynamics across the different stacking, the effective masses ($m^*$) extracted from parabolic fits around the band extrema in the $\Gamma$-M direction are reported in Table~\ref{tab:effective_mass}.
 The $m^*$s can be put in relation with the exciton binding energy and hence the optical response via models like the Wannier-Mott hydrogenoid model.
Most of the stackings show that the $m^*$ of the electron is larger than that of the hole, indicating that conduction band is flatter than the valence band, thus
the exciton dynamics is expected to be dominated by the limited mobility of the electronic part of the exciton.

\begin{figure}
  \centering
  \includegraphics[width=1.\columnwidth]{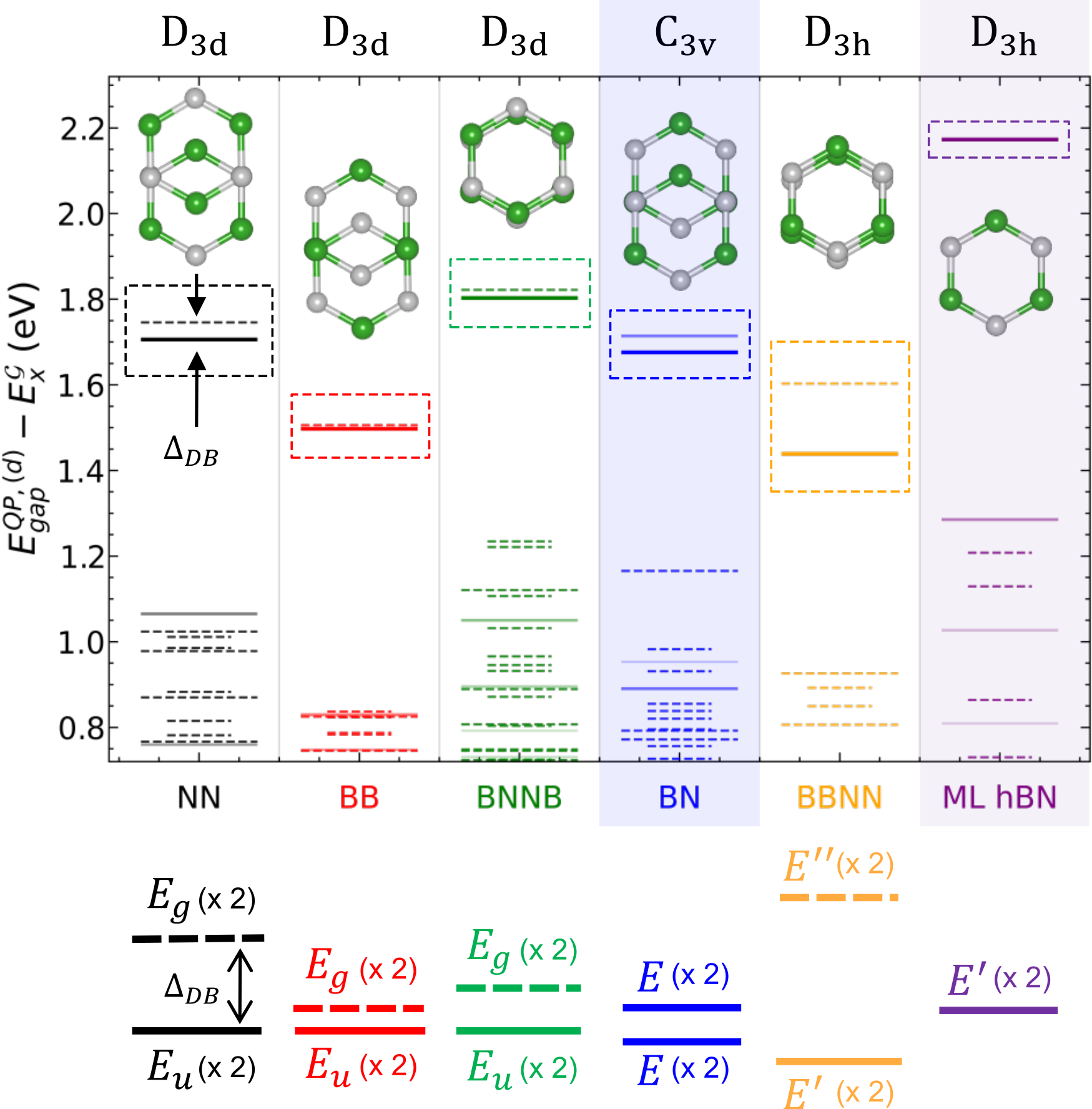}
  \caption{Lowest-lying exciton binding-energy levels. Solid and dashed lines denote bright and dark excitons, respectively, while the color opacity of the bright states is proportional to their oscillator strength. Shorter and longer segments denote singly or doubly degenerate levels. The lower panels show the symmetry classification of the lowest excitonic states with the label ($\times 2$) highlighting their twofold degeneracy.  }
  \label{fig:BSEEGLEVELS}
\end{figure}

\begin{figure*}
  \centering
  \includegraphics[width=0.9\textwidth]{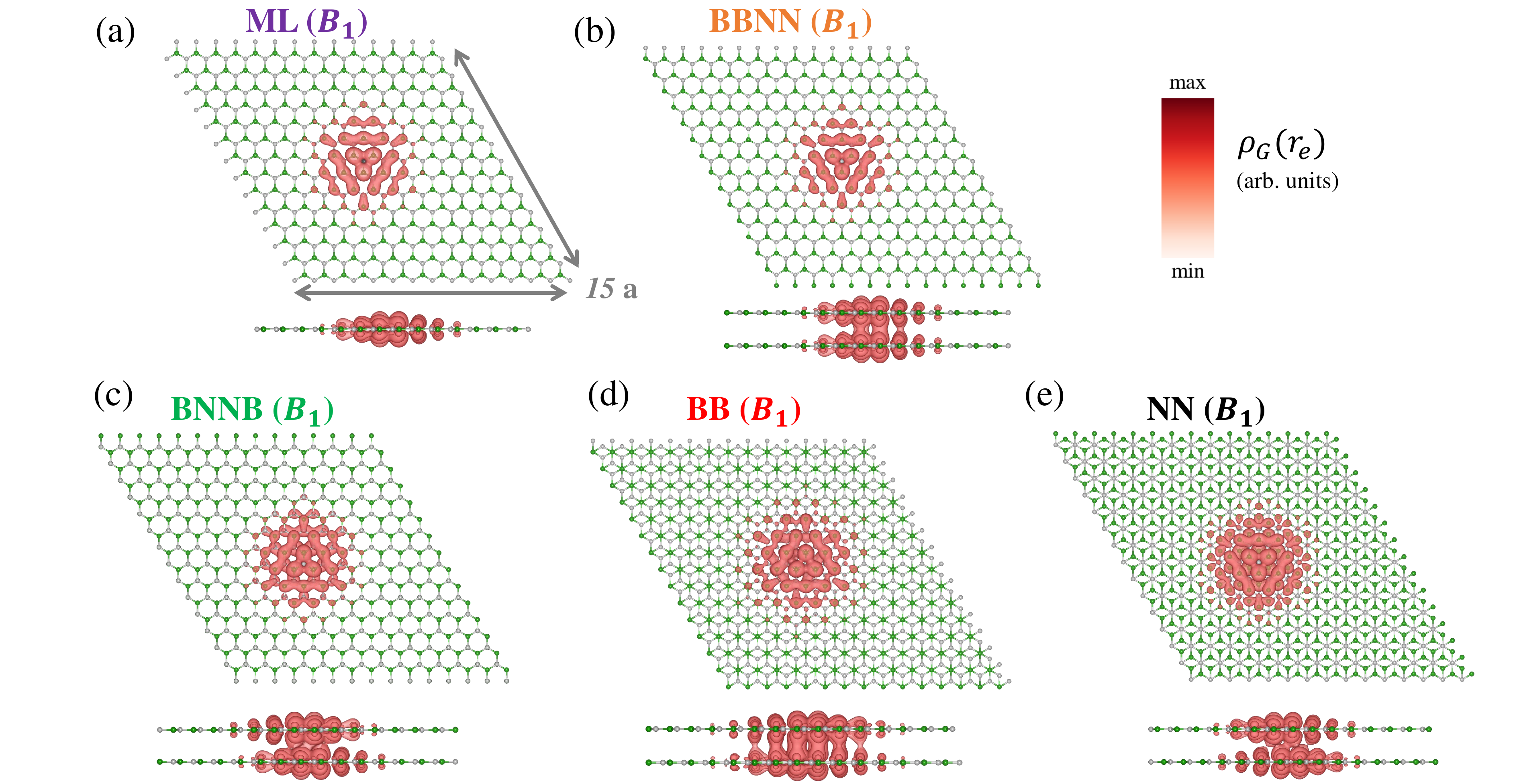}
  \caption{Real-space excitonic wave function of the first bright $B_1$ exciton for (a) ML, (b) BBNN, (c) BNNB, (d) BB, and (e) NN hBN. Top and side views are shown for each system. Each density is normalized independently for visualization.}
  \label{fig:exciton_real_space_wf}
\end{figure*}

\section{Ranking of low-energy excitons}
\label{APPEXC}

Figure~\ref{fig:BSEEGLEVELS} reports the lowest-lying BSE exciton binding energy levels. The solid lines correspond to bright excitonic states, with the color opacity proportional to their oscillator strength. The dashed lines represent dark states. The label  $\times 2$ indicates that the corresponding level is twofold degenerate; for clarity, the degenerate levels are shown with a extended line width. The lower panel illustrates the symmetry origin of the lowest Davydov exciton pair. In the noncentrosymmetric structures: the lowest exciton transforms as $E$ for BN with $C_{3v}$ symmetry, $E^{''}$ for BBNN with $D_{3h}$ symmetry, and $E^{'}$ for ML hBN with $D_{3h}$ symmetry. The $E$ and $E^{'}$ states are bright for in-plane polarization, whereas the $E^{''}$ state is dark.

\section{Real-space exciton wave functions}
\label{APPWFC}

Figure~\ref{fig:exciton_real_space_wf} reports the real-space excitonic wave function of the first bright $B_1$ exciton in ML- and BL-hBN for different stacking. First panel corresponds to the ML case, while the next panels represent the BL configurations (BBNN, BNNB, BB, NN). For each structure, the upper raw shows the top (in-plane) view, and the lower raw shows the side (lateral) view. The red isosurfaces illustrate the $\rho_{\mathcal{G}}(\mathbf{r}_e;\mathbf{r}_h^0)$ of the first bright exciton.

Additionally, in Fig.~\ref{fig:fixed_hole_CT_DD} we show the electron-density profiles along the out-of-plane direction, $P(z)$, of the lowest $D_1$ and $B_1$ excitons in BB, BNNB, NN, and BBNN. The hole is fixed near an N atom in either the top or bottom layer. Dashed and solid lines represent the dark and bright excitons, respectively. The insets show the corresponding exciton envelope wavefunction. For each stacking, the similar $P(z)$ profiles and momentum-space distributions show that $D_1$ and $B_1$ come from the same parent exciton, supporting their assignment as a DS pair. The main peaks show that these excitons are mostly on intralayer. The smaller peaks on the opposite layer are a finite CT contribution.  Across the stackings, the DS tends to increase as the difference in opposite-layer weight between $D_1$ and $B_1$ becomes larger. This trend is stronger in BBNN, where the strongest difference is accompanied by the largest splitting, $\Delta_{DB}=164$ meV.

%


%

\end{document}